\documentclass{llncs}
\usepackage{hyperref}
\newcommand{\myref}[1]{\hyperref[#1]{\ref*{#1}}}
\usepackage[noadjust]{cite}
\usepackage{amsmath}
\usepackage[caption=false,font=footnotesize]{subfig}
\usepackage{url}
\usepackage{graphicx}
\usepackage{amssymb}
\everymath{\displaystyle}
\usepackage{comment}

\usepackage[shortlabels]{enumitem}
\usepackage{slashbox}
\usepackage{textcomp}
\usepackage{gensymb}
\usepackage{hhline}

\DeclareMathOperator{\sign}{sign}
\newcommand{\mytranspose}[1]{{#1}^\top}

\begin{document}

\title{Approximate Fitting of a Circular Arc\\When Two Points Are Known}

\author{Alexander Gribov}

\institute
{
  Esri\\
  380 New York Street\\
  Redlands, CA 92373\\
  \email{agribov@esri.com}
}

\maketitle

\begin{abstract}
\boldmath
The task of approximating points with circular arcs is performed in many applications, such as polyline compression, noise filtering, and feature recognition. However, the development of algorithms that perform a significant amount of circular arcs fitting requires an efficient way of fitting circular arcs with complexity $O{\left( 1 \right)}$. The elegant solution to this task based on an eigenvector problem for a square nonsymmetrical matrix is described in \cite{FittingOfCircularArcsWithO1Complexity}. For the compression algorithm described in \cite{OptimalCompression}, it is necessary to solve this task when two points on the arc are known. This paper describes a different approach to efficiently fitting the arcs and solves the task when one or two points are known.
\end{abstract}

\section{Introduction}

The purpose of this paper is to solve fitting a circular arc to a set of points (or segments) with complexity $O{\left( 1 \right)}$ when two points on the arc and moments up to the fourth order are known.

In papers \cite{ThomasReference2} and \cite{IchokuReference3}, fitting a circle is done by finding a circle with center $(x_c, y_c)$ and radius $r$, which minimizes the next equation
\begin{equation}
\sum\limits_{i = 1}^n \left(\left(\left(x_i-x_c\right)^2+\left(y_i-y_c\right)^2\right)-r^2\right)^2
,
\label{CircularArcSquareFitting}
\end{equation}
where $(x_i, y_i)$ are $i$-th point, $i=\overline{1..n}$.

This formula minimizes the squared differences between squared distances from the circle center to the points and square of the radius. The solution is found by using only the moments of $(x_i, y_i)$ with complexity $O{\left( 1 \right)}$. However, this leads to bias in the estimation of parameters \cite[see pp. 368-370]{ThomasReference2}. Suppose that each point has been fitted with $\epsilon_i$ error. Substituting it in \eqref{CircularArcSquareFitting} gives
\begin{equation*}
  \sum\limits_{i = 1}^n \left(\left(r+\epsilon_i\right)^2-r^2\right)^2
  =
  \sum\limits_{i = 1}^n \left( 2 r \cdot \epsilon_i + \epsilon_i^2 \right)^2 = \sum\limits_{i = 1}^n \left( \epsilon_i^2 \left( 2 r + \epsilon_i \right)^2 \right).
\end{equation*}
Assuming that $\epsilon_i$ are small compared to radius $r$ and neglecting higher orders
\begin{equation}
  4 r^2 \sum\limits_{i = 1}^n \epsilon_i^2.
  \label{CircularArcSquareFittingApproximateError}
\end{equation}

\begin{figure} [htb]
\centering
  \includegraphics[width = 0.8\columnwidth, keepaspectratio]{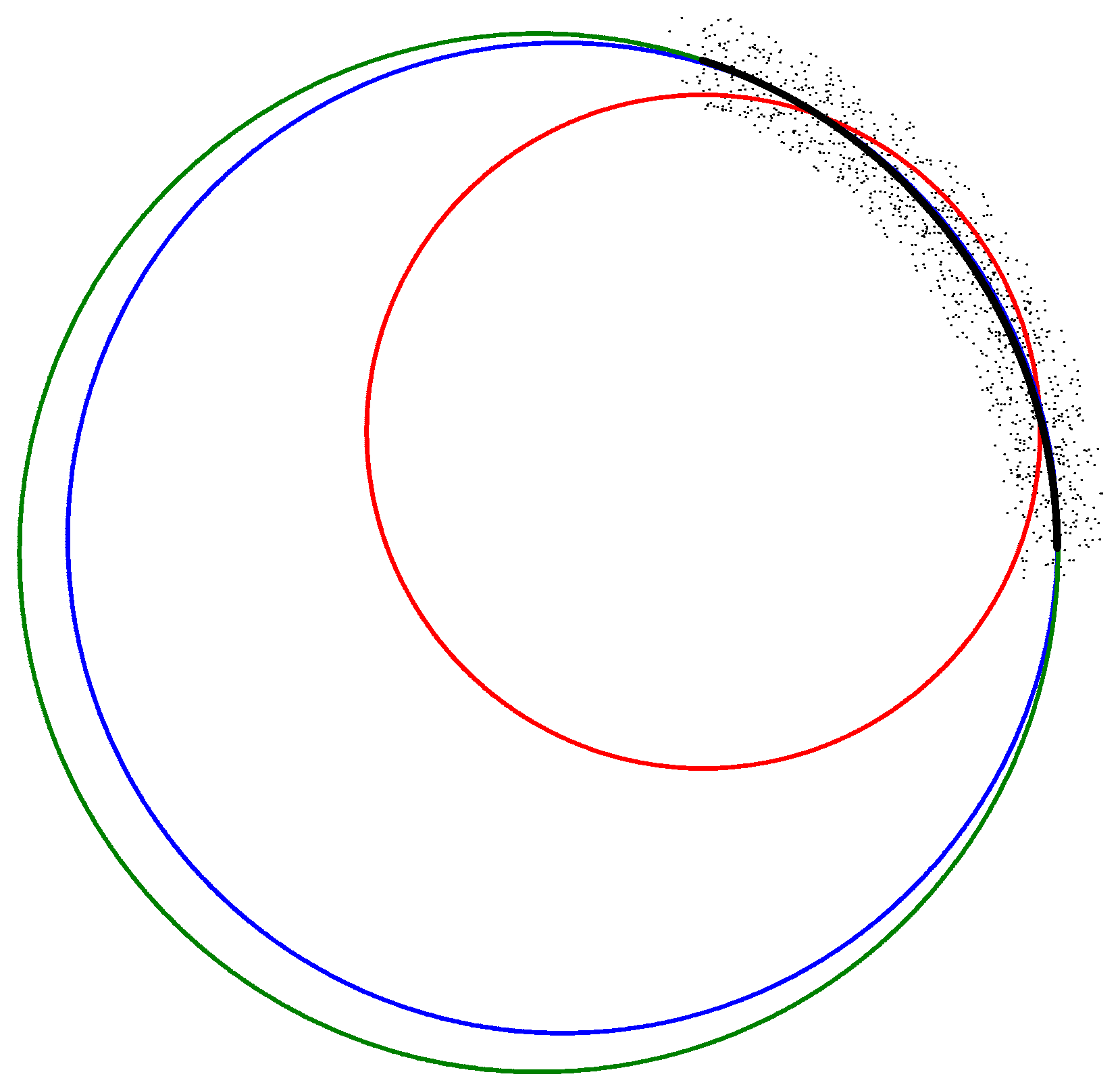}
  \caption
  {
    Comparison of fitting an arc using different approaches. The comparison is performed for the arc with $72\degree$, and uniform noise in the circle is proportional to $10$ percent of the arc radius. A total of $1,000$ random points were simulated along the arc with uniform steps. The black arc is a ground truth arc. The black dots are source points. The red circle is a solution based on fitting squares of distances (see~\eqref{CircularArcSquareFitting}). The green circle is a solution based on fitting distances (see~\eqref{CircularArcCorrectFitting}). The blue circle is a solution described in this paper (approximate solution of~\eqref{CircularArcFitting} found by one iteration of the algorithm described in \myref{AppendixMultiDimensionalOptimization}).
  }
  \label{fig:ApproximationOfCircle}
\end{figure}

From this formula, it is clear that the fitting is trying to decrease the radius to minimize \eqref{CircularArcSquareFitting}. When the points cover only a small part of a circle, the estimated center of the circle can move toward the arc to reduce the radius, see Fig.~\ref{fig:ApproximationOfCircle}. However, the errors are increased, while the overall penalty \eqref{CircularArcSquareFitting} is decreased. The smaller the angle of the arc, the worse the effect.

Minimizing the next equation was suggested in \cite{RobinsonReference6}:
\begin{equation}
\sum\limits_{i = 1}^n \left(\sqrt{\left(x_i-x_c\right)^2+\left(y_i-y_c\right)^2}-r\right)^2,
\label{CircularArcCorrectFitting}
\end{equation}
leading to the next error formula
\begin{equation}
\sum\limits_{i = 1}^n \epsilon_i^2.
\label{CircularArcFittingError}
\end{equation}

This doesn't encounter a problem like in \eqref{CircularArcSquareFittingApproximateError}. However, minimizing \eqref{CircularArcCorrectFitting} requires an iterative approach, which analyzes all points $(x_i, y_i)$ leading to an algorithm with complexity $O{\left( n \right)}$. The efficient algorithm to find the minimum of \eqref{CircularArcCorrectFitting} is described in \cite{PaperArcFitting}.

\section{Algorithm}
The solution to remove the square root from \eqref{CircularArcCorrectFitting} was developed in \cite{DirectLeastSquaresFittingOfAlgebraicSurfaces}, \cite[see p. 675]{SquareRootApproximation}, \cite{DirectLeastSquareFittingOfEllipses}, and \cite{FittingOfCircularArcsWithO1Complexity}. From \eqref{CircularArcSquareFittingApproximateError} comes an idea that dividing \eqref{CircularArcSquareFitting} by $4 r^2$ and minimizing it will produce a result closer to \eqref{CircularArcCorrectFitting} because it is close to \eqref{CircularArcFittingError}. The approximation based on the Taylor expansion of the square root by the first two terms gives exactly this solution. Approximation of $\sqrt{x}$ at $x=1$:
\begin{equation*}
\sqrt{x}\approx 1+\frac{1}{2}(x-1)+O\left(x^2\right)=\frac{1}{2}+\frac{x}{2}+O\left(x^2\right)
.
\end{equation*}

Applying this approximation to \eqref{CircularArcCorrectFitting}
\begin{multline}
\sum\limits_{i = 1}^n \left(\sqrt{\left(x_i-x_c\right)^2+\left(y_i-y_c\right)^2}-r\right)^2=\\
=r^2\sum\limits_{i = 1}^n\left(\sqrt{\frac{\left(x_i-x_c\right)^2+\left(y_i-y_c\right)^2}{r^2}}-1\right)^2\approx\\
\approx r^2\sum\limits_{i = 1}^n \left(\left(\frac{1}{2}+\frac{\left(x_i-x_c\right)^2+\left(y_i-y_c\right)^2}{2 r^2}\right)-1\right)^2=\\
=\frac{\sum\limits_{i = 1}^n \left(\left(\left(x_i-x_c\right)^2+\left(y_i-y_c\right)^2\right)-r^2\right)^2}{4 r^2}
.
\label{CircularArcFitting}
\end{multline}

Unlike \eqref{CircularArcCorrectFitting}, this formula can be minimized using only moments. The direct solution, based on conformal geometric algebra, is described in \cite{FittingOfCircularArcsWithO1Complexity}. The solution is based on finding eigenvalues of a square nonsymmetric matrix \cite[see (24)]{FittingOfCircularArcsWithO1Complexity}. This can be done using Schur factorization. The eigenvector corresponding to the minimal non-negative eigenvalue is the solution. Care should be taken in cases where the solution has close to zero eigenvalue due to round-off error.

In this paper, another approach to minimization of \eqref{CircularArcFitting} will be considered.

\section{Minimization of \eqref{CircularArcFitting}\label{SectionMinimizationOfEquationCircularArcFitting}}

Starting from a good estimate $\left( x_e, y_e, r_e \right)$, the search for the optimum can be performed in the next form:
$
\left(x_e + \Delta{x}, y_e + \Delta{y}, \sqrt{r_e ^ 2 + \Delta{x ^ 2} + \Delta{y ^ 2} + \Delta{r}} \right).
$

This covers all possible values of $\left(x_c,y_c,r\right)$ and gives the significant advantage for finding the minimum by removing the third and fourth order variables in the numerator of \eqref{CircularArcFitting}:
\begin{equation*}
f \left( \Delta{x}, \Delta{y}, \Delta{r} \right)
=
\frac
{
  \sum\limits_{i = 1}^n
  \left(
    \begin{aligned}
      x_i^2+y_i^2-
      2\left( x_e\cdot  x_i+ y_e\cdot  y_i\right)+
      \left(x_e^2+y_e^2-r_e^2\right)+\\
      +2\Delta{x} \left( x_e- x_i\right)+
      2\Delta{y} \left( y_e- y_i\right)-
      \Delta{r}
    \end{aligned}
  \right)^2
}
{
  4 \left( r_e ^ 2 + \Delta{x} ^ 2 + \Delta{y} ^ 2 + \Delta{r} \right)
}
.
\end{equation*}

Writing it from moments
\begin{equation}
  f \left( \Delta{x}, \Delta{y}, \Delta{r} \right)
  =
  \frac
  {
  \left(
  \begin{aligned}
    v+v_x\cdot \Delta{x}+v_y\cdot \Delta{y}+v_r\cdot \Delta{r}+\\
    +v_{x,x}\cdot \Delta{x}^2+v_{y,y}\cdot \Delta{y}^2+\Delta{r}^2+\\
    +v_{x,y}\cdot \Delta{x}\cdot \Delta{y}+v_{x,r}\cdot \Delta{x}\cdot \Delta{r}+v_{y,r}\cdot \Delta{y}\cdot \Delta{r}
  \end{aligned}
  \right)
  }
  {
    4 \left(r_e^2+\Delta{x}^2+\Delta{y}^2+\Delta{r}\right)
  }
  ,
  \label{EverythingFormula}
\end{equation}
where
\begin{equation*}
  \begin{aligned}
  v&=\left(M_{4,0}+2M_{2,2}+M_{0,4}\right)-4\left(M_{3,0}+M_{1,2}\right)x_e-4\left(M_{2,1}+M_{0,3}\right)y_e+\\
   &+8M_{1,1}\cdot x_e\cdot y_e+2M_{2,0}\cdot z_x+2M_{0,2}\cdot z_y-4\left(M_{1,0}\cdot x_e+M_{0,1}\cdot y_e\right)z+z^2,\\
  v&_x=4\left(-\left(M_{3,0}+M_{1,2}\right)+\left(3M_{2,0}+M_{0,2}\right)x_e+\right.\\
   &\left.+2M_{1,1}\cdot y_e-2M_{0,1}\cdot x_e\cdot y_e-M_{1,0}\cdot z_x+x_e\cdot z\right),\\
  v&_y=4\left(-\left(M_{2,1}+M_{0,3}\right)+\left(M_{2,0}+3M_{0,2}\right)y_e+\right.\\
   &\left.+2M_{1,1}\cdot x_e-2M_{1,0}\cdot x_e\cdot y_e-M_{0,1}\cdot z_y+y_e\cdot z\right),\\
  v&_r=-2 \left(M_{2,0} + M_{0,2} - 2\left(M_{1,0}\cdot x_e+M_{0,1}\cdot y_e\right)+z\right),\\
  v&_{x,x}=4\left(M_{2,0}-2M_{1,0}\cdot x_e+x_e^2\right),\quad
  v_{y,y}=4\left(M_{0,2}-2M_{0,1}\cdot y_e+y_e^2\right),\\
  v&_{x,y}=8\left(M_{1,1}-M_{0,1}\cdot x_e-M_{1,0}\cdot y_e+x_e\cdot y_e\right),\\
  v&_{x,r}=4\left(M_{1,0}-x_e\right),\quad
  v_{y,r}=4\left(M_{0,1}-y_e\right),\\
  z&=x_e^2+y_e^2-r_e^2,\quad
  z_x=3x_e^2+y_e^2-r_e^2,\quad
  z_y=x_e^2+3y_e^2-r_e^2,\\
  M&_{g,h}
  =
  \dfrac{1}{n}
  \sum\limits_{i = 1}^n\left( x_i ^ g \cdot y_i ^ h \right).
  \end{aligned}
\end{equation*}

Minimization of \eqref{EverythingFormula} can be done using the approach described in \myref{AppendixMultiDimensionalOptimization}. To use that approach, it is sufficient to know the matrix of second derivatives up to the constant
\begin{equation*}
\begin{pmatrix}
\frac{\partial^2 f}{\partial^2 \Delta{x}} & &\\
\frac{\partial^2 f}{\partial \Delta{x} \partial \Delta{y}} & \frac{\partial^2 f}{\partial^2 \Delta{y}} &\\
\frac{\partial^2 f}{\partial \Delta{x} \partial \Delta{r}} & \frac{\partial^2 f}{\partial \Delta{y} \partial \Delta{r}} & \frac{\partial^2 f}{\partial^2 \Delta{r}}
\end{pmatrix}
\sim
\begin{pmatrix}
d_{x,x} & &\\
d_{x,y} & d_{y,y} &\\
d_{x,r} & d_{y,r} & d_{r,r}
\end{pmatrix}
,
\end{equation*}
where
\begin{equation*}
\left.
\begin{aligned}
d_{x,x}&=-2\left(v-v_{x,x}\cdot r_e^2\right)\cdot r_e^2,\\
d_{y,y}&=-2\left(v-v_{y,y}\cdot r_e^2\right)\cdot r_e^2,\\
d_{y,r}&=\left(-v_y+v_{y,r}\cdot r_e^2\right)\cdot r_e^2,
\end{aligned}
\right.
\quad
\left.
\begin{aligned}
d_{x,y}&=v_{x,y}\cdot r_e^4,\\
d_{x,r}&=\left(-v_x+v_{x,r}\cdot r_e^2\right)\cdot r_e^2,\\
d_{r,r}&=2\left(v-v_r\cdot r_e^2+r_e^6\right),
\end{aligned}
\right.
\end{equation*}
and the equation for directional search by direction
$\left( \alpha _x, \alpha _y, \alpha _r \right)$
is
\begin{equation*}
\frac
{
  \left(
  \begin{aligned}
    v+\left(v_x\cdot \alpha _x+v_y\cdot \alpha _y+v_r\cdot \alpha _r\right)t+\\
    +\left(v_{x,x}\cdot \alpha _x^2+v_{y,y}\cdot \alpha _y^2+\alpha _r^2+v_{x,y}\cdot \alpha _x\cdot \alpha _y+\right.\\
    \left.+v_{x,r}\cdot \alpha _x\cdot \alpha _r+v_{y,r}\cdot \alpha _y\cdot \alpha _r\right)t^2
  \end{aligned}
  \right)
}
{
  4 \left(r_e^2+\alpha _r\cdot t+\left(\alpha _x^2+\alpha _y^2\right)t^2\right)
}.
\end{equation*}

Looking at the numerator of \eqref{CircularArcFitting}, it would be reasonable to take a solution of \eqref{CircularArcSquareFitting}, described in \cite{ThomasReference2}, as a starting point.

Only a few iterations are needed to converge beyond machine precision. Because this approach is only an approximation, there is no need for such precision. In practice, one iteration is sufficient to get a good approximation.

When estimation of the center is known, the best estimation of the radius can be easily found. However, it does not give any improvement in speed.

An approximation of the sum of squared deviations from the polyline to an arc with the center  $\left( x_e, y_e \right)$ and radius $r_e$ (see \eqref{CircularArcCorrectFitting}) is found from \eqref{EverythingFormula} by setting $\Delta{x}$, $\Delta{y}$, and $\Delta{r}$ to zero and multiplying by $n$.
\begin{equation}
  n
  \frac
  {
    v
  }
  {
    4 r_e^2
  }
  .
  \label{EquationPenaltyFunction}
\end{equation}

The algorithm described in \cite{FittingOfCircularArcsWithO1Complexity} has the advantage of finding the global optimum, while the algorithm described in this paper can find the local optimum. This is likely to happen when the arc is close to the line. Otherwise, the results are identical.

The advantage of the approach described in this paper is the ability to reduce the amount of calculation by approximating the solution.

I have implemented both approaches. \href{https://software.intel.com/en-us/intel-mkl}{Intel Math Kernel Library $11.2$} was used to solve the nonsymmetric eigenvector problem in \cite{FittingOfCircularArcsWithO1Complexity}. The approach described in this paper is several times faster. However, it is difficult to make a fair comparison due to the different ways of implementing and optimizing the code. When speed is not a concern, the approach described in \cite{FittingOfCircularArcsWithO1Complexity} is preferred.

\section{Evaluation of Fitting Quality and Speed}

A comparison of fitting quality is done in Table~\ref{tab:TableComparison}. There are situations when fitting cannot be done for all or some approaches. To overcome this difficulty, the median is used. The number of points is $1,000$. The number of simulations is $1,000,001$. The ground truth circle has a radius of $1$. The noise is uniform and perpendicular to the circumference of the circle.

\begin{table} [p]
\caption{Comparison of the fitting quality of different approaches depending on the angle of the approximated arc and noise level. In each cell, there are three rows with two numbers. The first row used the method described in this paper with one iteration, the second row used a complete optimization, and the third row used an iterative solution based on fitting distances \cite{PaperArcFitting} (see~\eqref{CircularArcCorrectFitting}). In each row, the first number is a median error in the estimation of the center; the second number is a median error in the radius.}
\centering
\begin{tabular}[c]{ r || c | c | c | c | c |}
 & \multicolumn{5}{c |}{Noise level}\\
\cline{2-6}
 & 1e-5 & 1e-4 & 1e-3 & 1e-2 & 1e-1\\
\hhline{=||=|=|=|=|=|}
\shortstack{1\degree \\ {} \\ {} \\ {}} & \shortstack{\\ 9.37e-3;9.37e-3\\9.37e-3;9.37e-3\\9.37e-3;9.37e-3} & \shortstack{9.26e-2;9.26e-2\\9.31e-2;9.31e-2\\9.31e-2;9.31e-2} & \shortstack{8.49e-1;7.40e-1\\6.36e-1;5.37e-1\\NaN;NaN} & \shortstack{1.00;9.90e-1\\1.00;9.90e-1\\NaN;NaN} & \shortstack{1.00;9.24e-1\\1.00;9.24e-1\\NaN;NaN}\\
\hline
\shortstack{2\degree \\ {} \\ {} \\ {}} & \shortstack{\\ 2.34e-3;2.34e-3\\2.34e-3;2.34e-3\\2.34e-3;2.34e-3} & \shortstack{2.34e-2;2.34e-2\\2.34e-2;2.34e-2\\2.34e-2;2.34e-2} & \shortstack{9.11e-1;9.11e-1\\2.26e-1;2.26e-1\\2.26e-1;2.26e-1} & \shortstack{9.17e-1;7.81e-1\\9.81e-1;9.44e-1\\NaN;NaN} & \shortstack{1.00;9.23e-1\\1.00;9.23e-1\\NaN;NaN}\\
\hline
\shortstack{3\degree \\ {} \\ {} \\ {}} & \shortstack{\\ 1.04e-3;1.04e-3\\1.04e-3;1.04e-3\\1.04e-3;1.04e-3} & \shortstack{1.04e-2;1.04e-2\\1.04e-2;1.04e-2\\1.04e-2;1.04e-2} & \shortstack{1.70e-1;1.70e-1\\1.04e-1;1.04e-1\\1.04e-1;1.04e-1} & \shortstack{7.40e-1;5.28e-1\\7.40e-1;6.10e-1\\NaN;NaN} & \shortstack{1.00;9.21e-1\\1.00;9.21e-1\\NaN;NaN}\\
\hline
\shortstack{4\degree \\ {} \\ {} \\ {}} & \shortstack{\\ 5.86e-4;5.86e-4\\5.86e-4;5.86e-4\\5.86e-4;5.86e-4} & \shortstack{5.86e-3;5.86e-3\\5.86e-3;5.86e-3\\5.86e-3;5.86e-3} & \shortstack{5.82e-2;5.82e-2\\5.85e-2;5.85e-2\\5.85e-2;5.85e-2} & \shortstack{8.26e-1;7.52e-1\\5.22e-1;4.86e-1\\NaN;NaN} & \shortstack{1.00;9.20e-1\\1.00;9.20e-1\\NaN;NaN}\\
\hline
\shortstack{5\degree \\ {} \\ {} \\ {}} & \shortstack{\\ 3.75e-4;3.75e-4\\3.75e-4;3.75e-4\\3.75e-4;3.75e-4} & \shortstack{3.75e-3;3.75e-3\\3.75e-3;3.75e-3\\3.75e-3;3.75e-3} & \shortstack{3.75e-2;3.75e-2\\3.75e-2;3.75e-2\\3.75e-2;3.75e-2} & \shortstack{7.17e-1;7.15e-1\\3.63e-1;3.61e-1\\NaN;NaN} & \shortstack{1.00;9.17e-1\\1.00;9.17e-1\\NaN;NaN}\\
\hline
\shortstack{10\degree \\ {} \\ {} \\ {}} & \shortstack{\\ 9.38e-5;9.37e-5\\9.38e-5;9.37e-5\\9.38e-5;9.37e-5} & \shortstack{9.38e-4;9.37e-4\\9.38e-4;9.37e-4\\9.38e-4;9.37e-4} & \shortstack{9.39e-3;9.37e-3\\9.39e-3;9.37e-3\\9.39e-3;9.37e-3} & \shortstack{4.66e-1;4.69e-1\\9.49e-2;9.47e-2\\9.48e-2;9.47e-2} & \shortstack{9.97e-1;9.01e-1\\9.98e-1;9.01e-1\\NaN;NaN}\\
\hline
\shortstack{20\degree \\ {} \\ {} \\ {}} & \shortstack{\\ 2.36e-5;2.34e-5\\2.36e-5;2.34e-5\\2.36e-5;2.34e-5} & \shortstack{2.36e-4;2.34e-4\\2.36e-4;2.34e-4\\2.36e-4;2.34e-4} & \shortstack{2.36e-3;2.34e-3\\2.36e-3;2.34e-3\\2.36e-3;2.34e-3} & \shortstack{2.36e-2;2.35e-2\\2.36e-2;2.35e-2\\2.37e-2;2.35e-2} & \shortstack{4.73e-1;4.64e-1\\7.61e-1;7.24e-1\\NaN;NaN}\\
\hline
\shortstack{30\degree \\ {} \\ {} \\ {}} & \shortstack{\\ 1.05e-5;1.04e-5\\1.05e-5;1.04e-5\\1.05e-5;1.04e-5} & \shortstack{1.05e-4;1.04e-4\\1.05e-4;1.04e-4\\1.05e-4;1.04e-4} & \shortstack{1.05e-3;1.04e-3\\1.05e-3;1.04e-3\\1.05e-3;1.04e-3} & \shortstack{1.05e-2;1.04e-2\\1.05e-2;1.04e-2\\1.05e-2;1.04e-2} & \shortstack{6.65e-1;6.63e-1\\1.79e-1;1.72e-1\\NaN;NaN}\\
\hline
\shortstack{60\degree \\ {} \\ {} \\ {}} & \shortstack{\\ 2.71e-6;2.53e-6\\2.71e-6;2.53e-6\\2.71e-6;2.53e-6} & \shortstack{2.71e-5;2.53e-5\\2.71e-5;2.53e-5\\2.71e-5;2.53e-5} & \shortstack{2.71e-4;2.53e-4\\2.71e-4;2.53e-4\\2.71e-4;2.53e-4} & \shortstack{2.71e-3;2.54e-3\\2.71e-3;2.54e-3\\2.71e-3;2.54e-3} & \shortstack{2.85e-2;2.61e-2\\2.92e-2;2.64e-2\\2.83e-2;2.64e-2}\\
\hline
\shortstack{90\degree \\ {} \\ {} \\ {}} & \shortstack{\\ 1.27e-6;1.09e-6\\1.27e-6;1.09e-6\\1.27e-6;1.09e-6} & \shortstack{1.27e-5;1.09e-5\\1.27e-5;1.09e-5\\1.27e-5;1.09e-5} & \shortstack{1.27e-4;1.09e-4\\1.27e-4;1.09e-4\\1.27e-4;1.09e-4} & \shortstack{1.27e-3;1.09e-3\\1.27e-3;1.09e-3\\1.27e-3;1.09e-3} & \shortstack{1.30e-2;1.12e-2\\1.30e-2;1.12e-2\\1.29e-2;1.11e-2}\\
\hline
\shortstack{180\degree \\ {} \\ {} \\ {}} & \shortstack{\\ 4.25e-7;2.44e-7\\4.25e-7;2.44e-7\\4.25e-7;2.44e-7} & \shortstack{4.25e-6;2.44e-6\\4.25e-6;2.44e-6\\4.25e-6;2.44e-6} & \shortstack{4.25e-5;2.45e-5\\4.25e-5;2.45e-5\\4.25e-5;2.45e-5} & \shortstack{4.25e-4;2.47e-4\\4.25e-4;2.47e-4\\4.25e-4;2.45e-4} & \shortstack{4.27e-3;4.92e-3\\4.27e-3;4.92e-3\\4.27e-3;2.59e-3}\\
\hline
\shortstack{270\degree \\ {} \\ {} \\ {}} & \shortstack{\\ 2.87e-7;1.21e-7\\2.87e-7;1.21e-7\\2.87e-7;1.21e-7} & \shortstack{2.87e-6;1.21e-6\\2.87e-6;1.21e-6\\2.87e-6;1.21e-6} & \shortstack{2.87e-5;1.21e-5\\2.87e-5;1.21e-5\\2.87e-5;1.21e-5} & \shortstack{2.87e-4;1.26e-4\\2.87e-4;1.26e-4\\2.87e-4;1.21e-4} & \shortstack{2.88e-3;4.96e-3\\2.88e-3;4.96e-3\\2.88e-3;1.53e-3}\\
\hline
\shortstack{360\degree \\ {} \\ {} \\ {}} & \shortstack{\\ 2.63e-7;1.07e-7\\2.63e-7;1.07e-7\\2.63e-7;1.07e-7} & \shortstack{2.63e-6;1.07e-6\\2.63e-6;1.07e-6\\2.63e-6;1.07e-6} & \shortstack{2.63e-5;1.07e-5\\2.63e-5;1.07e-5\\2.63e-5;1.07e-5} & \shortstack{2.63e-4;1.12e-4\\2.63e-4;1.12e-4\\2.63e-4;1.07e-4} & \shortstack{2.64e-3;4.96e-3\\2.64e-3;4.96e-3\\2.64e-3;1.43e-3}\\
\hline
\end{tabular}
\label{tab:TableComparison}
\end{table}

Assuming that the center of the unit circle is known and the errors are distributed uniformly in $\left[ -w, w \right]$, $0 \leq w \wedge w \leq 1$, perpendicular to the circumference of the unit circle, the bias is equal to
\begin{equation*}
  \dfrac{1}{2 w}
  \int\limits_{1 - w}^{1 + w}
  {
    \left(
      \left(
        \dfrac{1}{2}
        +
        \dfrac{x^2}{2}
      \right)
      -
      1
    \right)
    dx
  }
  =
  \dfrac{w^2}{6}
.
\end{equation*}
Therefore, the estimated radius tends to be larger than the true radius. For example, for the uniform noise of $ 7 $ percent, the error in the estimation of radius is less than $ 0.1 $ percent.

When enough information is available to reconstruct an arc, all approaches perform equally well. The approximation of the square root has minimum effect, unless the noise is large and the arc is small. When the arc is too small for the approach described in this paper, one iteration is definitely not enough. Note that in this case, the arc is very close to the line. The approach described in \cite{PaperArcFitting} sometimes fails because it might perform division by numbers close to zero; however, the arc can be reconstructed using other approaches. It is possible to improve the stability of convergence in \cite{PaperArcFitting} by providing a better starting point, for example, by the algorithm described in this paper, with a sufficient number of iterations.

The comparison of fitting speed is performed by averaging the time used by each approach for $10,000$~simulations of the $72\degree$ arc with uniform noise proportional to $10$~percent of the arc radius. Points are simulated along the arc with uniform steps. When moments up to the fourth order are known, the described approach becomes faster at $5$~points. The described approach has constant complexity, while the iterative approach has linear complexity. For $100$~points, the advantage in speed is about $9$~times. When moments have to be calculated, it is faster at $6$~points. For $100$~points, the advantage in speed is about $3.5$~times.

\section{Optimal Arc When One Point Is Known}

For some tasks, one point on the arc is known in advance. The arc should pass through that point. Knowing the position of the center determines the radius: $r = \sqrt{\left( \left( x_c - x_a \right) ^2 + \left( y_c - y_a \right) ^2 \right)}$, where $\left( x_a , y_a \right)$ is a point on the arc. Substituting this into \eqref{CircularArcFitting}
\begin{equation}
\frac
{
  \sum\limits_{i = 1}^n 
  \left(
    \left(
      \left( x_i - x_c \right)^2 + \left( y_i - y_c \right)^2
    \right)-
    \left( \left( x_c - x_a \right) ^2 + \left( y_c - y_a \right) ^2 \right)
  \right) ^ 2
}
{
  4 \left( \left( x_c - x_a \right) ^2 + \left( y_c - y_a \right) ^2 \right)
}
.
\label{eq:OnePointFitting}
\end{equation}

The solution described in Sect.~\myref{SectionMinimizationOfEquationCircularArcFitting} can be applied. The differences are that the optimization is performed in two-dimensional space, and the initial solution can be found from the least squares approach.

Multiplying the numerator and denominator of \eqref{eq:OnePointFitting} by $s^2$ and replacing $x_c \cdot s$ and $y_c \cdot s$ by $u_x$ and $u_y$, respectively, and setting
$
  u
  =
  \mytranspose
  {
  \begin{pmatrix}
    u_x &
    u_y &
    s
  \end{pmatrix}
  }
$
\begin{equation}
  \dfrac
  {
    \mytranspose{u} A \; u
  }
  {
    4 \; \mytranspose{u} B \; u
  }
  ,
  \label{eq:MatrixRatioForFittingArcThroughOnePoint}
\end{equation}
where
\begin{equation*}
  A
  =
  \begin{pmatrix}
    a_{x,x} & a_{x,y} & a_{x,1}\\
    a_{x,y} & a_{y,y} & a_{y,1}\\
    a_{x,1} & a_{y,1} & a_{1,1}
  \end{pmatrix}
  ,
  \quad
  B
  =
  \begin{pmatrix}
    1 & 0 & -x_a\\
    0 & 1 & -y_a\\
    -x_a & -y_a & x^2_a + y^2_a
  \end{pmatrix}
  ,
\end{equation*}
\begin{equation*}
  \begin{aligned}
  a_{x,x} = & 4 \left( M_{2,0} - 2 M_{1,0} \cdot x_a + x^2_a \right),\quad
  a_{y,y} = 4 \left( M_{0,2} - 2 M_{0,1} \cdot y_a + y^2_a \right),\\
  a_{1,1} = & M_{4,0} + 2 M_{2,2} + M_{0,4} - 2 \left( M_{2,0} + M_{0,2} \right) \left( x^2_a + y^2_a \right) + \left( x^2_a + y^2_a \right)^2,\\
  a_{x,y} = & 4 \left( M_{1,1} - M_{1,0} \cdot y_a - M_{0,1} \cdot x_a + x_a y_a \right),\\
  a_{x,1} = & -2 \left( M_{3,0} + M_{1,2} - \left( M_{2,0} + M_{0,2} \right) x_a - M_{1,0} \left( x^2_a + y^2_a \right) + \left( x^2_a + y^2_a \right) x_a \right),\\
  a_{y,1} = & -2 \left( M_{2,1} + M_{0,3} - \left( M_{2,0} + M_{0,2} \right) y_a - M_{0,1} \left( x^2_a + y^2_a \right) + \left( x^2_a + y^2_a \right) y_a \right).
  \end{aligned}
\end{equation*}

Equation \eqref{eq:MatrixRatioForFittingArcThroughOnePoint} is a generalized Rayleigh quotient. Note that $A$ and $B$ are symmetric non-negative matrices. The solution can be found by the generalized singular value decomposition of $\sqrt{A}$ and $\sqrt{B}$. Square root matrices can be found from singular value decompositions such as
\begin{equation}
    \sqrt{A} = L_A^{\frac{1}{2}} \mytranspose{X}_A,\quad
    \sqrt{B} = L_B^{\frac{1}{2}} \mytranspose{X}_B,
  \label{eq:SquareRootAB}
\end{equation}
where
$L_A$ and $L_B$ are eigenvalue matrices of $A$ and $B$, respectively; $X_A$ and $X_B$ are eigenvector matrices of $A$ and $B$, respectively.

From generalized singular value decomposition for $\sqrt{A}$ and $\sqrt{B}$ follows
\begin{equation}
    \sqrt{A} = U D_A \left( 0, R \right) \mytranspose{Q},\quad
    \sqrt{B} = V D_B \left( 0, R \right) \mytranspose{Q},
  \label{eq:GeneralizedSingularValueDecomposition}
\end{equation}
where
$U$, $V$, and $Q$ are orthogonal matrices; $R$ is an upper triangular matrix.

From \eqref{eq:SquareRootAB} and \eqref{eq:GeneralizedSingularValueDecomposition} follows
\begin{equation*}
    A = Q \mytranspose{\left( 0, R \right)} D_A^2 \left( 0, R \right) \mytranspose{Q},\quad
    B = Q \mytranspose{\left( 0, R \right)} D_B^2 \left( 0, R \right) \mytranspose{Q}.
\end{equation*}

The smallest ratio of squares of eigenvalues is the solution. The center of the arc is recovered from $u$ as $\left( \dfrac{u_x}{s}, \dfrac{u_y}{s} \right)$.

\section{Optimal Arc When Two Points Are Known\label{OptimalArcWhenTwoPointsAreKnown}}

There are tasks when the starting and ending points of the arc are known (or any two points lying on the arc). Therefore, the center of the arc should lie on some line of $\left( x_c , y_c \right) = \left( x_p + \alpha \cdot t , y_p + \beta \cdot t \right)$, where $\left( x_p , y_p \right)$ is a point on the line, $\left( \alpha , \beta \right)$ is the direction of the line ($\alpha ^2 + \beta ^ 2 = 1$), and $t$ is any value. Knowing the position of the center determines the radius:
\begin{equation*}
  r = \sqrt{\left( x_a - \left( x_p + \alpha \cdot t \right) \right) ^2 + \left( y_a - \left( y_p + \beta \cdot t \right) \right) ^2},
\end{equation*}
where $\left( x_a , y_a \right)$ is one of the points on the arc. Substituting this in \eqref{CircularArcFitting}
\begin{multline}
    f \left( t \right)
    =
    \frac
    {
      \sum\limits_{i = 1}^n 
      \left(
        \begin{aligned}
          \left(
            \left( x_i - \left( x_p + \alpha \cdot t \right) \right)^2 + \left( y_i - \left( y_p + \beta \cdot t \right) \right)^2
          \right) -\\
          - \left( \left( x_a - \left( x_p + \alpha \cdot t \right) \right) ^2 + \left( y_a - \left( y_p + \beta \cdot t \right) \right) ^2 \right)
        \end{aligned}
      \right) ^ 2
    }
    {
      4 \left( \left( x_a - \left( x_p + \alpha \cdot t \right) \right) ^2 + \left( y_a - \left( y_p + \beta \cdot t \right) \right) ^2 \right)
    }
    =
    \\
    =
    \frac
    {
      a_0 + a_1 \cdot t + a_2 \cdot t ^ 2
    }
    {
      b_0 + b_1 \cdot t + b_2 \cdot t ^ 2
    }
    ,
\label{EquationTwoPoints}
\end{multline}
where
\begin{equation*}
  \begin{aligned}
    a&_0 = q + x_p \cdot q_x + y_p \cdot q_y + x_p^2 \cdot q_{xx} + x_p \cdot y_p \cdot q_{xy} + y_p^2 \cdot q_{yy},\\
    a&_1 = \alpha \cdot q_x + \beta \cdot q_y + 2 \left( x_p \cdot \alpha \cdot q_{xx} + y_p \cdot \beta \cdot q_{yy} \right) + \left( x_p \cdot \beta + y_p \cdot \alpha \right) q_{xy},\\
    a&_2 = \alpha^2 \cdot q_{xx} + \alpha \cdot \beta \cdot q_{xy} + \beta^2 \cdot q_{yy},\\
    b&_0 = \left( x_a^2 + y_a^2 \right) - 2 \left( x_p \cdot x_a + y_p \cdot y_a \right) + \left( x_p^2 + y_p^2 \right),\\
    b&_1 = -2 \left( \alpha \cdot x_a + \beta \cdot y_a \right) + 2 \left( x_p \cdot \alpha + y_p \cdot \beta \right),\quad
    b_2 = 1,\\
    q& = \dfrac{1}{4} \left( \left( x_a^2 + y_a^2 \right)^2 - 2 \left( x_a^2 + y_a^2 \right) \left( M_{2,0} + M_{0,2} \right) + \left( M_{4,0} + 2 M_{2,2} + M_{0,4} \right) \right),\\
    q&_x = -x_a^3 + \left( x_a^2 + y_a^2 \right) M_{1,0} + x_a \left( \left( M_{2,0} + M_{0,2} \right) - y_a^2 \right) - \left( M_{3,0} + M_{1,2} \right),\\
    q&_y = -y_a^3 + \left( x_a^2 + y_a^2 \right) M_{0,1} + y_a \left( \left( M_{2,0} + M_{0,2} \right) - x_a^2 \right) - \left( M_{0,3} + M_{2,1} \right),\\
    q&_{xx} = x_a^2 - 2 x_a \cdot M_{1,0} + M_{2,0},\quad
    q_{yy} = y_a^2 - 2 y_a \cdot M_{0,1} + M_{0,2},\\
    q&_{xy} = 2 \left( x_a \cdot y_a - \left( x_a \cdot M_{0,1} + y_a \cdot M_{1,0} \right) + M_{1,1} \right).
  \end{aligned}
\end{equation*}

The minimum of~\eqref{EquationTwoPoints} can be easily found because it has the form of~\eqref{RatioEquation} in \myref{AppendixQuadratic}. An iterative algorithm is not needed to solve this problem.

The algorithm described in \cite{OptimalCompression} finds an optimal polyline within the tolerance of the source polyline, with the minimum number of vertices, and among them, with the minimum sum of the squared deviations from the optimal polyline. Extending this algorithm to support arcs requires efficient fitting of the arc from the known start and end points and evaluation of the sum of the squared deviations from the source polyline to an arc. An approximate solution \eqref{EquationPenaltyFunction} can be used instead of direct evaluation \eqref{CircularArcCorrectFitting}.

\section{Example: Recovering Arcs in a Cadastral Dataset\label{sec:Example}}

The approach described in this paper for efficiently fitting circular arcs is used in a compression algorithm, when vertices of the source polylines are not allowed to move. The algorithm minimizes the weighted number of segments (with penalty~2) and arcs (with penalty~3) while satisfying tolerance restrictions. Among all possible solutions, the solution with the minimum sum of squared deviations is chosen. A dynamic programming approach was used to find the optimal solution, see \cite{OptimalCompressionWithArcs}, \cite{PolylineGeneralizationCombinatorical}, \cite{DynamicCompressionWithArcs}, and \cite{OptimalCompression}.\footnote
{
  The penalty function in \cite{DynamicCompressionWithArcs} is a combination of perceptual and fitting errors. The perceptual error is
  $
    \delta \cdot \sin{\dfrac{\alpha}{2}}
  $,
  where
  $\delta$ is the segmentation penalty and $\alpha$ is the angle between adjacent segments. This gives preference to solutions with acute angles.
}

Fitting of arcs in \cite{OptimalCompressionWithArcs} was performed by checking tolerance when starting and ending vertices are fixed. It has the advantage of always finding an arc within tolerance; however, the computational complexity for each fitting is $O{\left( n \log{ \left( n \right) } \right)}$. This paper uses approximation to least squares fitting with complexity $O{\left( 1 \right)}$ described in Sect.~\myref{OptimalArcWhenTwoPointsAreKnown}. Although checking for the tolerance and proper sequence (zigzag) (\cite{OptimalCompressionWithArcs}, \cite{PaperArcFitting}, and \cite{OptimalCompression}) has complexity $O{\left( n \right)}$, it is only performed for optimal fits.

\begin{figure} [htb]
\centering
\includegraphics[width = \columnwidth, keepaspectratio]{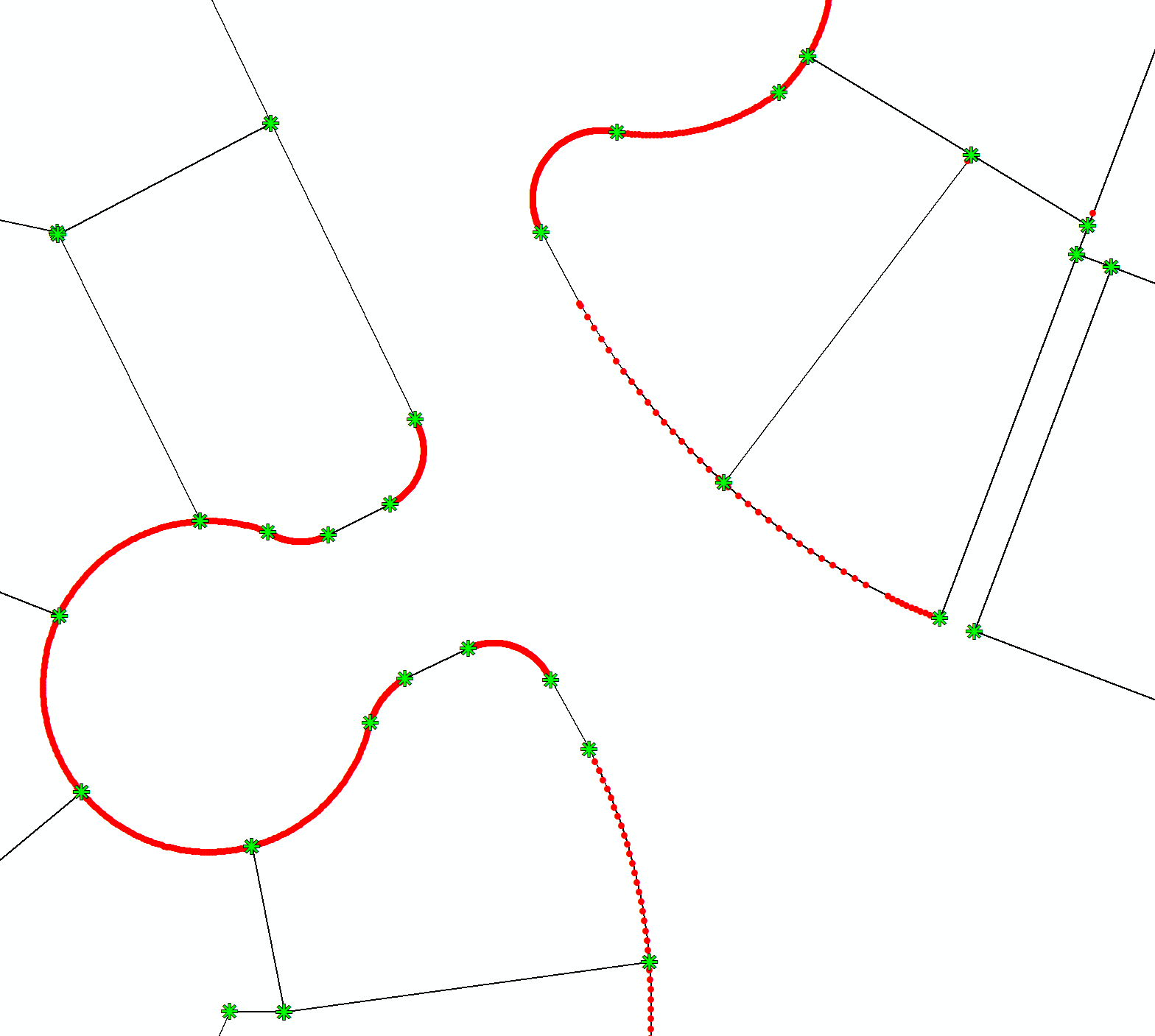}
\caption
{
  Part of a parcel map with lost circular arcs. A compression algorithm was applied to this data. The black lines are the source polylines, the red circles are vertices of the source polylines, and the green asterisks are resultant vertices. All original arcs were reconstructed.
}
\label{fig:ExampleCompressionWithArcs}
\end{figure}

An example is shown in Fig.~\ref{fig:ExampleCompressionWithArcs}. The original arcs were lost due to digitization, limitations of the format, projection, and so forth. The restoration of arcs is an important task because restoring original arcs creates cleaner databases and simplifies future editing.

\section{Conclusion}

This paper describes an efficient method of approximate fitting circular arcs. While all formulas are for a two-dimensional case, the algorithm can be generalized for higher dimensions (for example, fitting a sphere to points).

The direct solution to fit arcs is described in \cite{FittingOfCircularArcsWithO1Complexity}. This paper extends the solution to cases when one or two points on the arc are known.

Because the solution is based on fourth orders, it has a negative impact on the precision of calculations. This can be solved by shifting data to the origin of a coordinate system and/or using floating point numbers with a larger mantissa. Another solution is to place points into an integer coordinate system and calculate all moments using exact arithmetic on integer numbers. Than for fitting an arc to a subset of points, recalculate moments for the origin closer to the center of an arc.

There is no evaluation of how well the fit is done. An additional algorithm is necessary to perform this check, as described in \cite[see Sect.~3]{PaperArcFitting}.

\section*{Acknowledgment}

The author would like to thank Linda Thomas, Lois Stuart, and Mary Anne Chan for proofreading this paper.

\clearpage

\appendix

\renewcommand{\thesection}{Appendix~\Roman{section}}

\section{Finding the Global Minimum of the Ratio of Quadratic Equations\label{AppendixQuadratic}}

\begin{equation}
\frac{a_0 + a_1 \cdot x + a_2 \cdot x ^ 2}{b_0 + b_1 \cdot x + b_2 \cdot x ^ 2},
\label{RatioEquation}
\end{equation}
where $a_i$ and $b_i$ are known coefficients $i=\overline{0..2}$.

Coefficients should satisfy
\begin{equation}
\forall{x}, a_0 + a_1 \cdot x + a_2 \cdot x ^ 2 \geq 0.
\label{RestrictionNotNegativeNumerator}
\end{equation}

Two cases will be analyzed separately:
\begin{enumerate}[label = \emph{\arabic*}., ref = \emph{\arabic*}]
\item $b_2 \neq 0$.
\label{enum: GeneralCaseSolution}

The domain will be restricted to
\begin{equation}
Q = \left\{ b_0 + b_1 \cdot x + b_2 \cdot x ^ 2 > 0 \right\}.
\label{RatioEquationDomain}
\end{equation}

Notice that \eqref{RatioEquation} has the same limits when $x \to - \infty$ and $x \to + \infty$.

The first derivative of \eqref{RatioEquation} equals
\begin{equation}
\frac
{
  \left(a_1 \cdot b_0 - a_0 \cdot b_1 \right) + 2 \left( a_2 \cdot b_0 - a_0 \cdot b_2 \right) \cdot x + \left( a_2 \cdot b_1 - a_1 \cdot b_2 \right) \cdot x ^ 2
}
{
  \left( b_0 + b_1 \cdot x + b_2 \cdot x ^ 2 \right) ^ 2
}.
\label{RatioDerivative}
\end{equation}

From \eqref{RatioEquationDomain}, it follows that the denominator \eqref{RatioDerivative} is always positive in $Q$. Therefore, it is sufficient to work with the numerator:
\begin{equation}
c_0 + c_1 \cdot x + c_2 \cdot x ^ 2,
\label{RatioDerivativeNumerator}
\end{equation}
where
$c_0 = a_1 \cdot b_0 - a_0 \cdot b_1$, $c_1 = 2 \left( a_2 \cdot b_0 - a_0 \cdot b_2 \right)$, and $c_2 = a_2 \cdot b_1 - a_1 \cdot b_2$.

From \eqref{RestrictionNotNegativeNumerator}, it follows that \eqref{RatioEquation} is not negative in $Q$. If the denominator of \eqref{RatioEquation} has real roots, then \eqref{RatioEquation}, when $x$ is approaching any root, goes to $+ \infty$ in $Q$ and $- \infty$ in the complement of $Q$ excluding roots (see example in Fig.~\ref{fig:ExampleRatioOfQuadratic}). Local extrema are found from roots of \eqref{RatioDerivativeNumerator} (Fig.~\ref{fig:ExampleDerivativeRatioOfQuadratic}). There is a special case, when in \eqref{RatioEquation} the numerator is equal to zero at one of the roots of the denominator. In this case, \eqref{RatioEquation} simplifies to the ratio of linear equations and doesn't have any global minimum.

\begin{figure} [htb]
  \centering
    \includegraphics[width = \columnwidth, keepaspectratio]{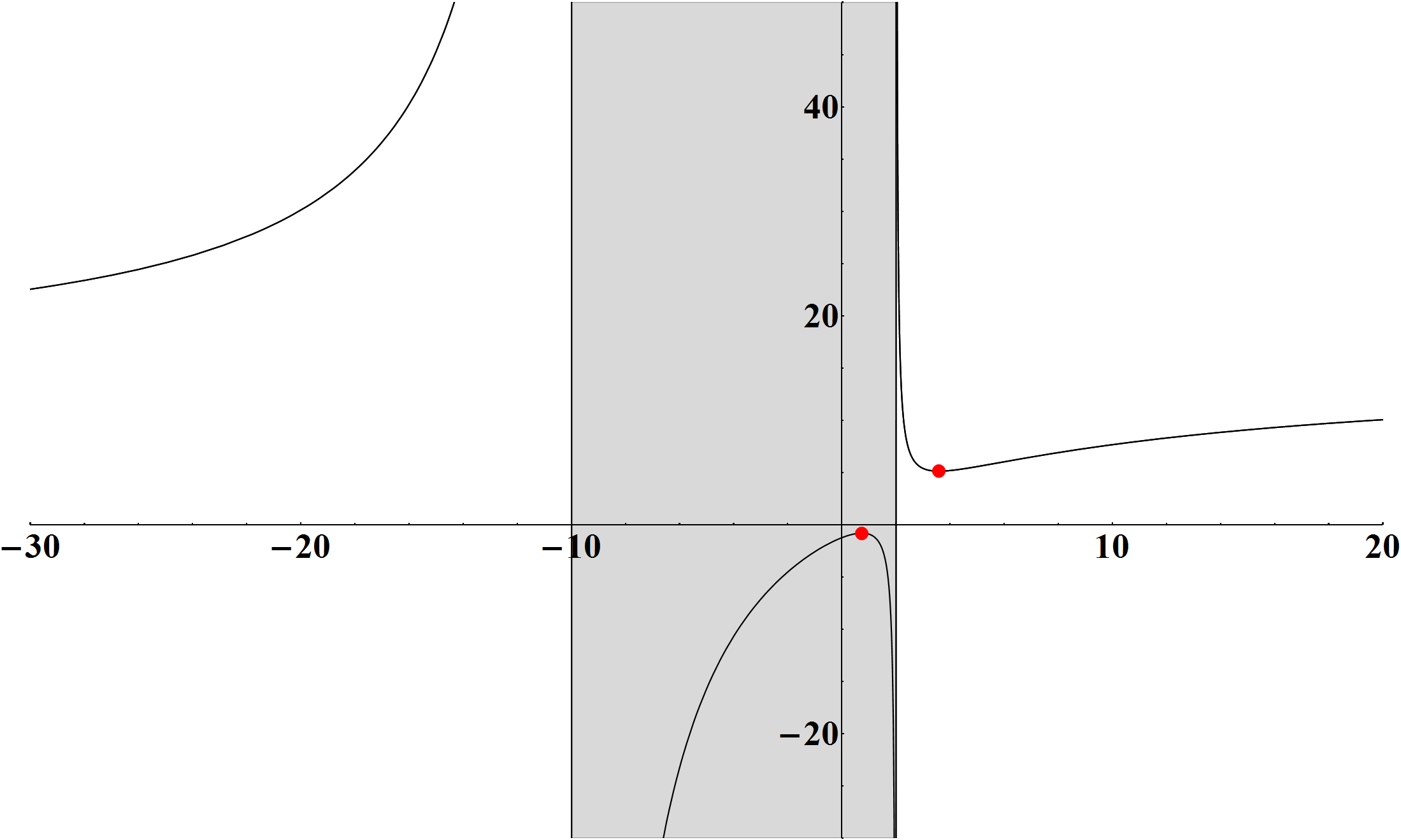}
  \caption
  {
    Example of \eqref{RatioEquation}. The area outside domain $Q$ is shown in gray. Local extrema are shown by red circles found as the solution of \eqref{RatioDerivativeNumerator} (see Fig.~\ref{fig:ExampleDerivativeRatioOfQuadratic}).
  }
  \label{fig:ExampleRatioOfQuadratic}
\end{figure}

\begin{figure} [!htb]
  \centering
    \includegraphics[width = \columnwidth, keepaspectratio]{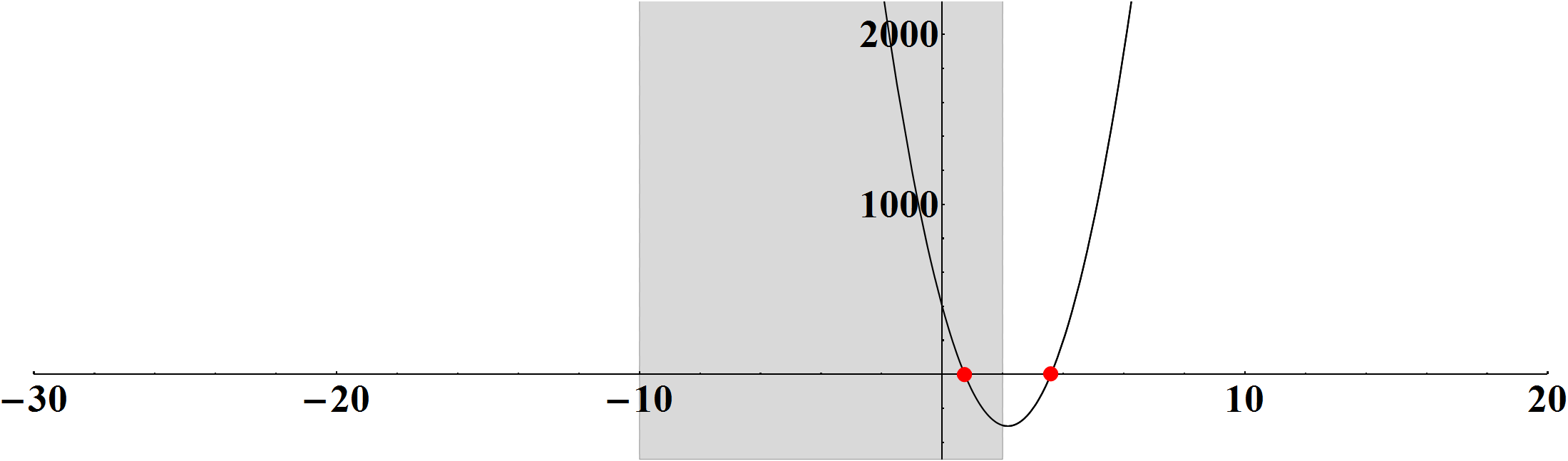}
  \caption
  {
    Example of \eqref{RatioDerivativeNumerator} corresponding to the function shown in Fig.~\ref{fig:ExampleRatioOfQuadratic}. Roots are shown by red circles.
  }
  \label{fig:ExampleDerivativeRatioOfQuadratic}
\end{figure}

The global minimum can be found from roots of the quadratic equation \eqref{RatioDerivativeNumerator}:
\begin{enumerate}[label = \emph{\alph*}.]
  \item If $c_2 > 0$ and the largest root of \eqref{RatioDerivativeNumerator} belongs to $Q$, then it is a global minimum.
  \item If $c_2 < 0$ and the smallest root of \eqref{RatioDerivativeNumerator} belongs to $Q$, then it is a global minimum.
  \item If $c_2 = 0$, $c_1 > 0$ and the single root of \eqref{RatioDerivativeNumerator} belongs to $Q$, then it is a global minimum.
  \item Otherwise, no global minimum exists.
\end{enumerate}

To summarize, the global minimum of \eqref{RatioEquation} can be found
\begin{equation}
\begin{cases}
-\frac{c_0}{c_1} & \mbox{if } c_2=0 \wedge c_1>0,\\
\frac{\sqrt{D}-c_1}{2 c_2} & \mbox{if } c_2 \neq 0 \wedge D>0 \wedge c_1 < 0,\\
\sign{\left( c_2 \right)} \cdot \sqrt{-\frac{c_0}{c_2}} & \mbox{if } c_2 \neq 0 \wedge D>0 \wedge c_1 = 0,\\
-\frac{2 c_0}{\sqrt{D}+c_1} & \mbox{if } c_2 \neq 0 \wedge D>0 \wedge c_1 > 0,\\
\mbox{no solution} & \mbox{otherwise},
\label{RatioFinalSolution}
\end{cases}
\end{equation}
where $D=c_1^2-4 c_0 \cdot c_2$ is discriminant of \eqref{RatioDerivativeNumerator} if the value is inside $Q$.

\item $b_2 = 0$. It is sufficient to evaluate the solution of the next equation to show that this case can be properly solved by \ref{enum: GeneralCaseSolution}:
$
\frac{a_0+a_1\cdot x+a_2\cdot x^2}{x}
$,
where $a_i$ are known coefficients $i=\overline{0..2}$.
The domain will be restricted to
$
Q = \left\{ x > 0 \right\}.
$
The first derivative multiplied by $x ^ 2$ equals
$
-a_0 + a_2 \cdot x ^ 2.
$

From that global minimum
\begin{equation}
\begin{cases}
\sqrt{\frac{a_0}{a_2}} & \mbox{if } a_0 \cdot a_2 > 0,\\
\mbox{no solution} & \mbox{otherwise},
\end{cases}
\label{RatioXFinalSolutionPartialCase}
\end{equation}

Notice that solution \eqref{RatioXFinalSolutionPartialCase} is equal to solution \eqref{RatioFinalSolution}. Therefore, it is sufficient to use \eqref{RatioFinalSolution} for both cases.

Another way to prove that the solution for the case \ref{enum: GeneralCaseSolution} gives the proper solution (when $b_2 = 0$) is to consider $\lim_{b_2 \to 0}$ of \eqref{RatioFinalSolution}.
\end{enumerate}


\section{Minimization of Multidimensional Function~$\operatorname{f}{\left( x \right)}$, $x \in \mathbb{R} ^ n$\label{AppendixMultiDimensionalOptimization}}

Suppose the minimum of $\operatorname{f} \left( x \right)$ along any direction can be found. Assume that the second derivatives can also be found.

The next algorithm is suggested:
\begin{enumerate}[label = \emph{\alph*}., ref = \emph{\alph*}]
  \item Let $i = 0$. Define the starting point $x_0$.
  \item \label{IterativeAlgorithmSecondDerivativeStep} Find the second derivative matrix at $x_{i}$, and find all eigenvectors.
  \item For each eigenvector, from $x_{i}$ point, search along the eigenvector direction for minimum $x_{i + 1}$. Set $i = i + 1$. Because the number of eigenvectors is $n$, this step increases the index of $x$ by $n$.
  \item If $x_{i}$ is close to the minimum with enough precision (for example, by comparing with the previous estimate $x_{i - n}$), then stop; otherwise, go to step~\ref{IterativeAlgorithmSecondDerivativeStep}.
\end{enumerate}

In the case of quadratic functions, this algorithm converges to the minimum in one iteration consisting of searching from any starting point by $n$ direction.

\clearpage

\newcommand{\doi}[1]{\textsc{doi}: \href{http://dx.doi.org/#1}{\nolinkurl{#1}}}


\begingroup
\raggedright
\bibliographystyle{IEEEtran}
\bibliography{ApproximateFittingOfACircularArcWhenTwoPointsAreKnown}
\endgroup


\end{document}